# Measuring the Occupation-Level Impact of AbbVie Intelligence: AI Applicability Analysis, 2024–2025

**John Regan[1], Jon Stevens[1], and Brian Martin[1]**

[1]AbbVie, BTS Information Research, Enterprise Product Incubator for Capabilities (EPIC)

This paper presents an empirical analysis of AbbVie Intelligence's measurable impact on employee work activities across 192 distinct occupations in 2024 and 2025. Drawing on 598,744 deidentified AI conversations classified according to the O*NET Intermediate Work Activity (IWA) taxonomy, we compute occupation-level AI Applicability Scores that quantify the extent to which AI tools can meaningfully assist or automate real work at scale. Three convergent analyses are conducted: (1) longitudinal year-over-year trends from 2024 to 2025, (2) a quasi-experimental pre-post evaluation of the AbbVie Intelligence 3.0 platform release in August 2025, and (3) a pre-post evaluation of the AbbVie AI Learning Summit held in November 2025. Results demonstrate statistically significant improvements across all three dimensions. Mean AI Applicability Scores rose substantially from 2024 to 2025; the platform release produced a +10.0% gain ($p < 0.001$); and the AI Learning Summit produced a +6.68% gain ($p < 0.001$). These findings establish that both technological platform enhancements and structured enterprise AI education programs independently and substantially expand the reach of AI across the AbbVie workforce.

## Introduction

Artificial intelligence tools are increasingly embedded in enterprise workflows, transforming how employees across a wide range of job functions perform their day-to-day tasks. AbbVie Intelligence is AbbVie's internal AI platform, designed to assist employee productivity by applying large language model capabilities to a broad spectrum of work activities. As the platform continues to evolve, quantifying the measurable impact of product releases on workforce applicability is essential for evidence-based investment decisions, change management strategy, and workforce planning.

A growing body of academic and industry research has established rigorous methods for measuring AI's relevance to occupational activities at scale [1,2,3,8,10]. Studies leveraging conversations from commercial AI platforms—including Microsoft Copilot and Anthropic's Claude—demonstrate that generative AI is applicable to substantial fractions of knowledge-work tasks across diverse occupational categories [3,8,10]. These external benchmarks make a compelling case for enterprise-level measurement, yet internal analyses of proprietary AI platforms within specific organizational contexts remain limited in the published literature. AbbVie's breadth of occupations—spanning pharmaceutical research, clinical operations, regulatory affairs, technology, and corporate functions—provides a uniquely rich environment in which to examine the heterogeneous impact of AI tools across a modern, research-intensive enterprise.

This paper addresses three primary research questions:

1. How have AI Applicability Scores across AbbVie occupations evolved from 2024 to 2025, and which roles and work activities have experienced the greatest change?
2. Did the release of AbbVie Intelligence 3.0 in August 2025 produce a measurable, statistically significant improvement in occupation-level AI Applicability Scores?
3. Did the AbbVie AI Learning Summit in November 2025 translate into measurable improvements in how employees apply AI to their work, and how does this educational effect compare to platform-driven gains?

## Data Collection

Our analysis draws on usage data from AbbVie Intelligence collected throughout 2024 and 2025. The data analyzed was comprised of:

- 192 distinct occupations
- 307 Intermediate Work Activities (IWAs) from the O*NET taxonomy
- 598,744 total AI conversations (293,134 from 2024 and 305,610 from 2025)

Each conversation was classified according to the work activities it addressed, following the methodology established in the Microsoft Research paper on AI applicability to occupations [3]. Conversations were sampled randomly and thus are representative of AbbVie Intelligence usage. Data privacy guardrails were upheld by running LLM classifiers over deidentified conversation logs that were scrubbed of any personally identifiable information (PII).

## O*NET Taxonomy Framework

We employ the O*NET (Occupational Information Network) taxonomy, a comprehensive framework maintained by the U.S. Department of Labor that provides a hierarchical classification of work:

- Generalized Work Activities (GWAs): Broad categories of work
- Intermediate Work Activities (IWAs): 332 specific work tasks that span multiple occupations
- Occupations: Specific job roles (e.g., "Biochemists and Biophysicists")
- Tasks: Detailed activities within specific occupations

This taxonomy enables systematic mapping of AI conversations to standardized work activities, facilitating comparison across different occupations and industries. O*NET provides detailed descriptions of the knowledge, skills, abilities, and tasks required for a wide range of jobs. This framework is crucial for occupational assessments because it offers standardized, up-to-date information that helps employers, job seekers, and HR professionals evaluate job requirements, match candidates to roles, and support workforce planning and career development. In addition to Microsoft, studies by Anthropic have used this framework to conduct similar AI impact analyses.

## AI Applicability Score

The primary outcome variable in this study is the AI Applicability Score, a composite metric computed at the occupation level. It quantifies the extent to which non-trivial AI usage can successfully complete activities corresponding to significant portions of an occupation's tasks. Conceptually, the score captures two dimensions of AI impact:

- Augmentation (User Goal): AI assists employees in completing tasks more effectively.
- Automation (AI Action): AI completes tasks with minimal or no human intervention.

Scores are derived from data collected through employee interactions with the AbbVie Intelligence platform. The score incorporates weighted activity coverage, task completion rates, and scope assessments from both user and AI perspectives.

*Figure 1: AI Applicability Score Formula*

$$a_i^{user} = \sum_{j \in \mathbf{IWA}_i} w_{ij} 1[f_j \geq 0.0005] C_j S_j^{user}$$

$$a_i^{AI} = \sum_{j \in \mathbf{IWA}_i} w_{ij} 1[f_j \geq 0.0005] C_j S_j^{AI}$$

$$a_i = \frac{a_i^{user} + a_i^{AI}}{2}$$

- $i$ = occupation (O*NET SOC code)
- $j$ = Intermediate Work Activity (IWA)
- $w_{ij}$ = importance & relevance-weighted fraction of work in occupation $i$ composed of IWA $j$
- $f_j$ = activity share (frequency threshold: 0.05%)
- $C_j$ = task completion rate for IWA $j$
- $S_j^{user}$ = scope_assistance (fraction where user-side scope is "moderate" or higher)

# Comparison of 2024 Scores with 2025

Of the 150 occupations scored in 2024, 147 were also present in the 2025 dataset, providing a stable longitudinal cohort for direct comparison. An additional 42 occupations entered the analysis in 2025, reflecting expanded workforce coverage, while 3 occupations were retired.

As illustrated in Figure 2, the 2024 score distribution was approximately symmetric around its mean of 0.175 and showed no occupations exceeding a score of 0.40. By 2025, this profile had changed fundamentally, with seven occupations crossing the 0.40 threshold for the first time. Meanwhile, the proportion of occupations in the low-applicability band (<0.10) remained relatively stable at 22 (2024) vs. 25 (2025), indicating that the uplift was concentrated in the middle and upper ranges, and not simply a uniform upward shift.

*Figure 2: AI Applicability Score Distribution*

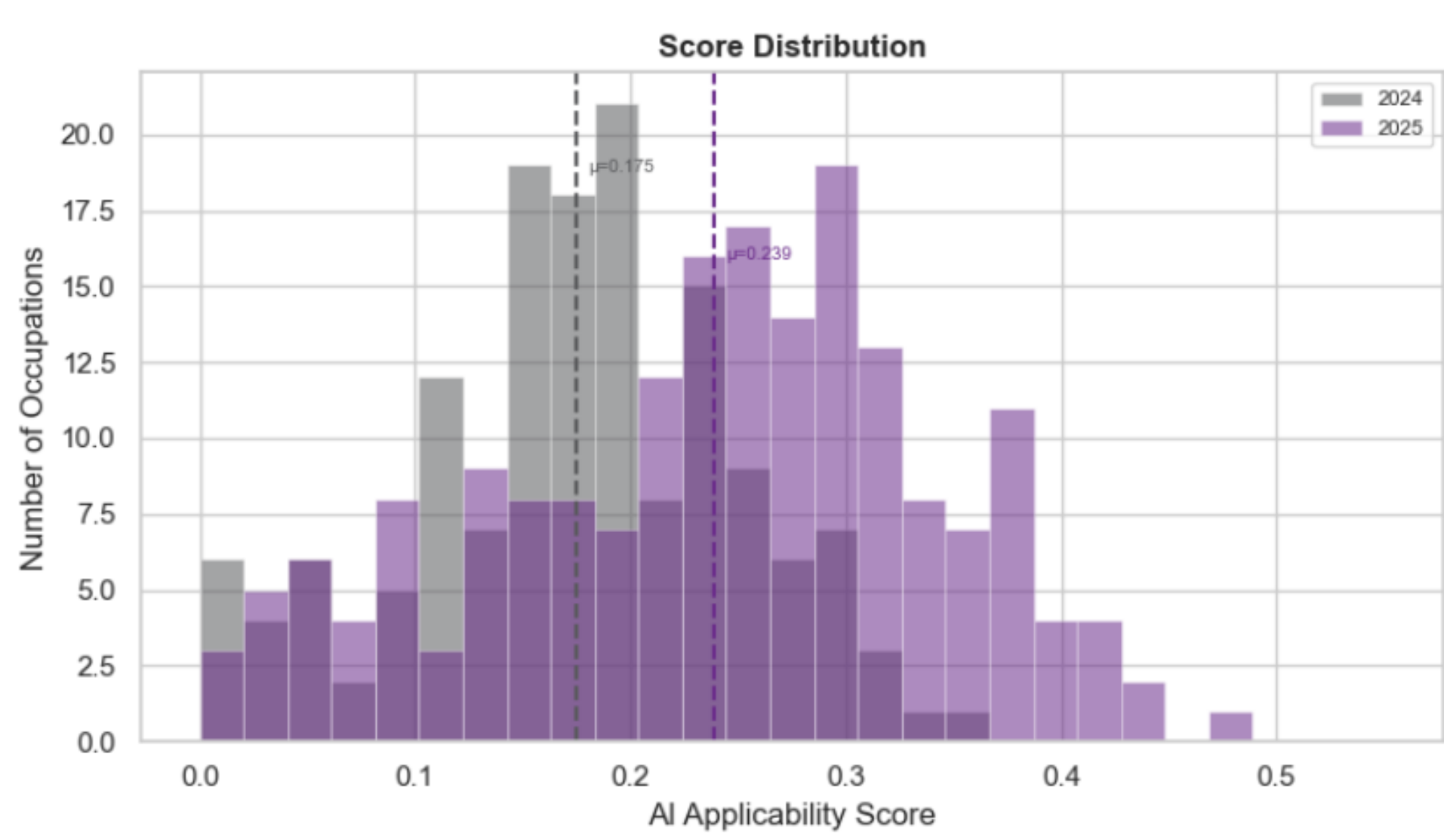

In 2025, the composition of occupations demonstrated a notable shift. Non-clinical and technology roles gained traction in AI usage.  This suggests broader adoption across operational and technical functions beyond traditional clinical settings.

*Figure 3: Top Occupation Improvements in AI Applicability Score, 2024 to 2025*

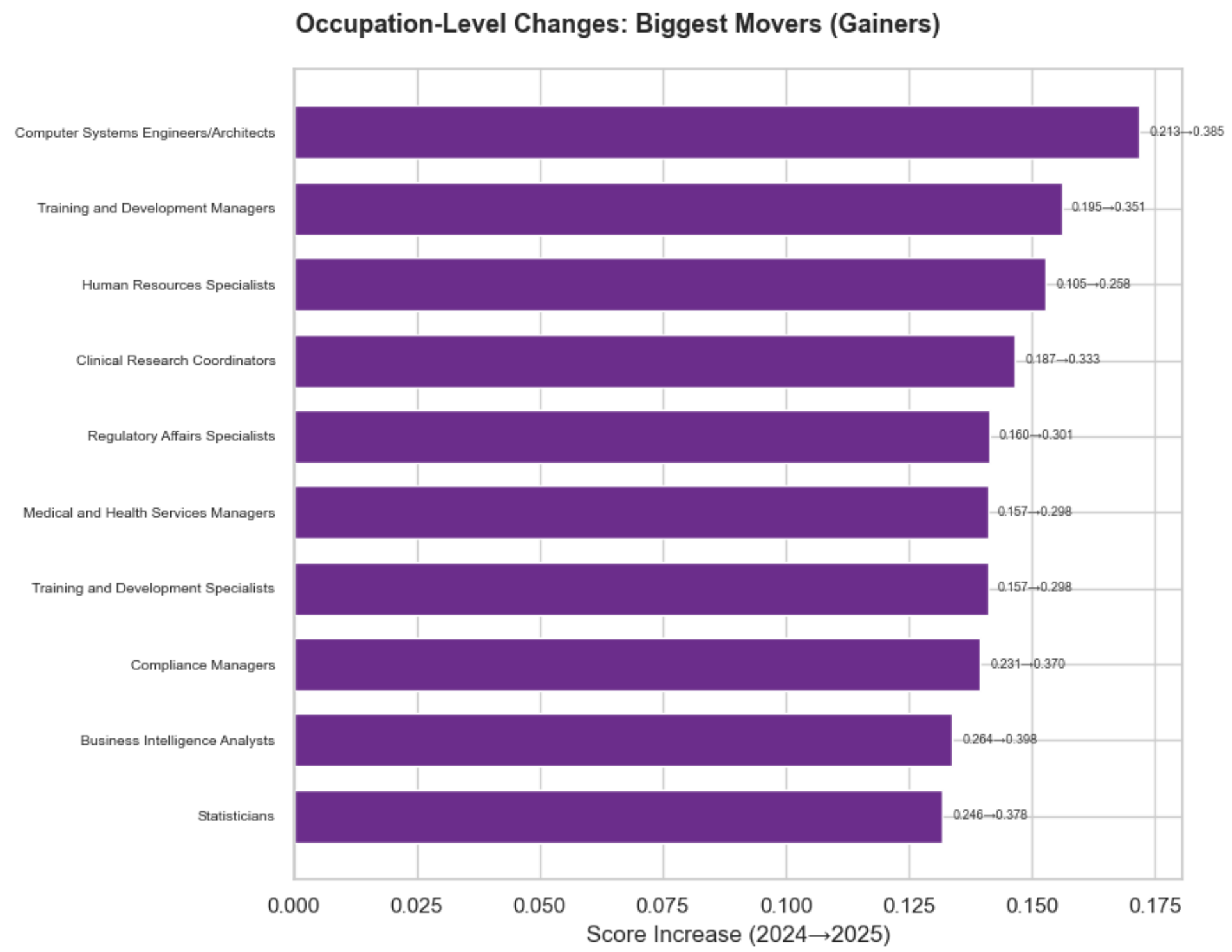

Below are the top 10 roles that with the largest AI Applicability score increases detailed from Figure 3.

- **Computer Systems Engineers/Architects (+0.172):** Computer Systems Engineers and Architects used AI for a wide range of hands-on technical tasks. The most common were software and infrastructure topics: writing and debugging code (React, Python, JavaScript, SCSS, VBA, SQL), configuring cloud and container platforms, implementing authentication flows, troubleshooting system issues, and working with data platforms. Database and data tasks were also frequent. There were also notable instances of users seeking conceptual explanations of technical topics.
- **Training and Development Managers (+0.156):** Training and Development Managers used AI primarily to develop and refine instructional and organizational content. This included designing learning objectives for workshops, creating structured onboarding programs, drafting training announcements and survey questions, summarizing post-training feedback, building presentation decks and executive summaries, and conceptualizing gamified or themed training experiences.
- **Human Resources Specialists (+0.153):** Human Resources Specialists most frequently used AI to draft, refine, and translate professional communications across a wide range of HR activities. These included onboarding announcements, rejection and offer emails, interview scheduling, performance feedback, payroll and benefits clarifications, recognition messages, and internal newsletters. HR-specific operational tasks also appeared regularly: building hiring flowcharts, creating recruitment status updates, advising on disability and Medicare processes, conceptualizing team-building events, and producing LinkedIn posts for employer branding. The data shows a workforce straddling administrative, analytical, and strategic HR responsibilities that turns to AI for both writing and analytical support.

- **Clinical Research Coordinators (+0.146):** On the operational side, users sought help drafting and polishing emails covering site engagement, recruitment status, protocol deviations, monitoring visit observations, reimbursement logistics, budget negotiations, ICF revisions, and study start-up coordination.[1]
- **Regulatory Affairs Specialists (+0.141):** Regulatory Affairs Specialists used AI predominantly for editing and translating regulatory documents and communications. Many conversations involved polishing professional messages about submission timelines, agency interactions (EMA, MHRA, PMDA, MFDS, Health Canada), variation types, and PV reporting obligations. A strong secondary theme was regulatory knowledge support: users asked for explanations of pharmacovigilance (PV) aggregate reporting (PSUR vs. PBRER, ICH E2C), agency submission timelines, compassionate/individual patient use distinctions, SmPC content decisions, RMP updates, MedDRA coding guidance, ICH guidelines for impurity specifications, and GCP audit findings.
- **Medical and Health Services Managers (+0.141):** Many Medical and Health Services Managers used AI extensively for clinical site coordination, Clinical Research Associate (CRA) assignments, Medical Science Liaison (MSL) engagement, cross-functional meeting invitations, speaker event logistics, and internal status reporting. A secondary cluster involved medical content and terminology: requesting definitions of clinical acronyms (e.g. TEAE, EAER, CDAI), explaining statistical concepts, researching disease areas, and asking about pharmacovigilance challenges.
- **Training and Development Specialists (+0.141):** Training and Development Specialists used AI most often for writing and refining professional communications across training-related workflows. These included drafting emails to encourage survey participation, announcing new training materials and onboarding folders, composing ERG welcome messages, polishing recognition notes, rewording training objectives and sales differentiation messaging, and creating reminders for overdue training completions. Other tasks focused on instructional content development: building RACI charts for training programs, creating onboarding checklists and flowcharts, drafting storyboards for safety incident training, writing survey instructions, designing questionnaires, and developing developmental goal statements.
- **Compliance Managers (+0.139):** Compliance Managers' AI conversations were dominated by professional writing assistance — drafting, editing, and polishing business communications including emails, compliance risk statements, meeting invites, investigation summaries, handover updates, and pharmacovigilance documentation. A secondary cluster involved regulatory and quality compliance work: reviewing supplier audit questionnaires, analyzing pharmacovigilance system migrations, applying root cause analysis, interpreting GDPR concepts, classifying promotional message wording, and summarizing exception reports.
- **Business Intelligence Analysts (+0.134):** AI conversations among Business Intelligence Analysts centered heavily on data and analytics work, including SQL and database troubleshooting (Hive SQL, Oracle PL/SQL, Excel formula errors), Power BI development (DAX formulas, dashboard building, forecasting), and working with platforms like Cloudera, Smartsheet, and Foundry. A significant share also involved synthesizing business metrics — crafting Key Business Questions, defining sales and market metrics in plain language, and converting analytics insights into execution plans.
- **Statisticians (+0.132):** Statisticians used AI primarily for hands-on statistical programming and data manipulation, with R and SAS as the dominant tools. Recurring tasks included writing and debugging R code, SAS code, understanding statistical concepts (multicollinearity, sensitivity analysis, interim analysis boundaries, Bayesian designs), and navigating clinical trial data standards.

---

[1] Clinical research is just one of many areas within R&D represented in our data. AbbVie Intelligence was first advertised to the R&D function when it was released, leading to a predictable R&D "advantage" of 18.8% above the mean AI Applicability Score in the 2024 data. We had expected this advantage to diminish in the 2025 data, as the tool had been in use across all functions that entire year, but interestingly, the advantage grew to 25.0%.

In 2024, ‘Edit written Materials or documents’ was the dominant IWA. Tasks such as email editing, grammar correction, and translation comprised a majority of AI requests. However, advances in LLM capabilities coupled with organization-wide upskilling led to more expansive use cases and a significant shift/increase in other IWA tasks as shown in Figure 4.

*Figure 4: Top 10 IWA Work Share Shifts, 2024 to 2025*

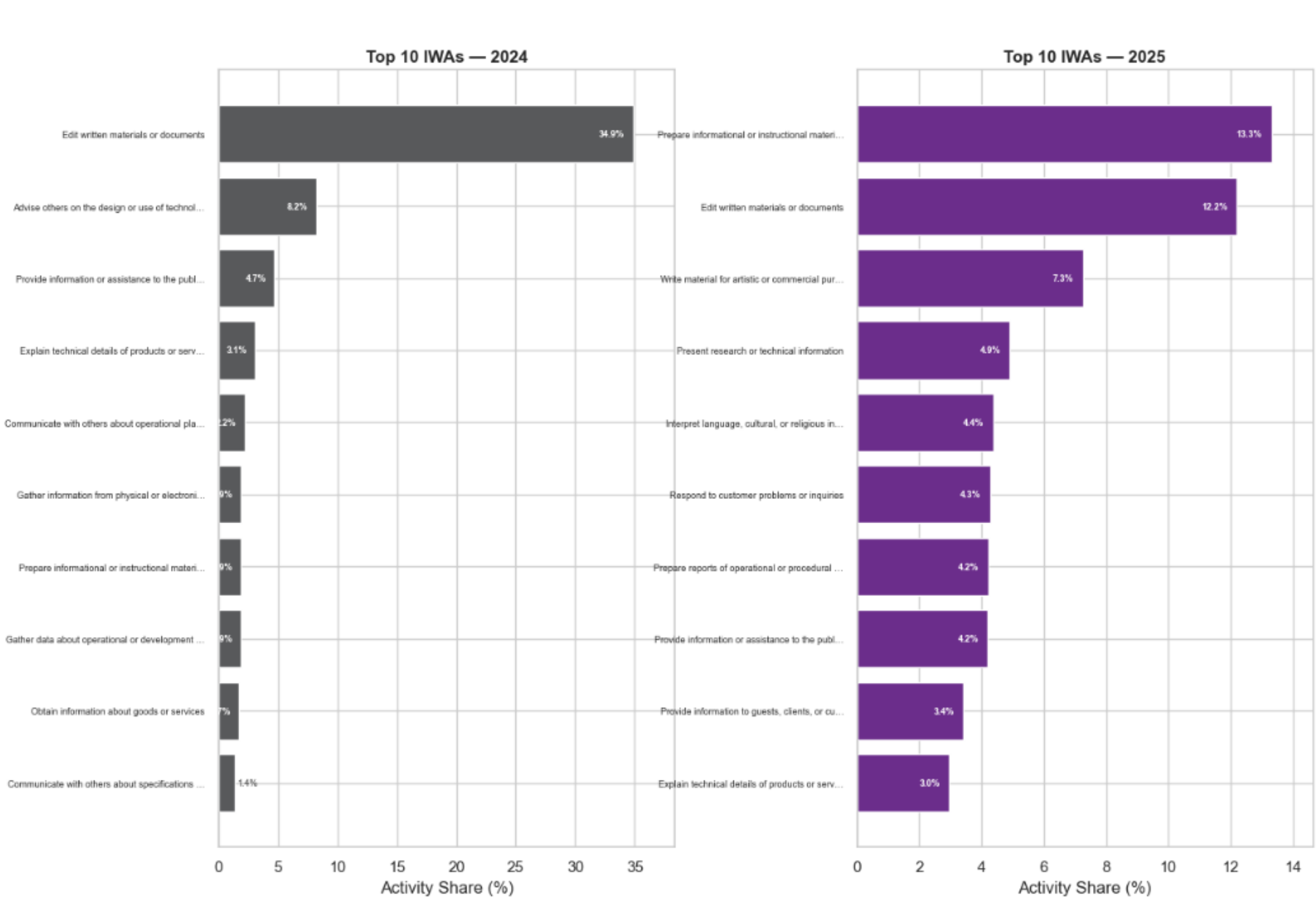

The top IWAs in 2025 are as follows:

- **Prepare informational or instructional materials:** Across this IWA, employees are primarily using AI to draft, refine, and structure professional communications — emails, announcements, training summaries, speaking notes, and meeting invites — spanning a wide range of business functions. Common subject areas include clinical trial coordination, call center procedures, regulatory communications, marketing/product announcements, HR and organizational updates, and cross-functional stakeholder outreach. Users frequently ask the AI to polish rough drafts, convert informal notes into formal text, or summarize lengthy documents into digestible formats. Other recurring tasks include drafting reward/recognition announcements, producing promotional headlines, and structuring instructional content. The AI's role is predominantly drafting and editing, with some summarization and light research support.
- **Edit written materials or documents:** This is one of the highest-volume IWAs, and a dominant use case across the tool. The overwhelming majority of interactions involve users submitting rough drafts — from single sentences to multi-paragraph emails — and asking the AI to improve grammar, tone, clarity, or professionalism. Content spans virtually every business domain: clinical trial monitoring communications, regulatory submissions, procurement and contracting, HR performance summaries, IT project updates, scientific manuscripts, financial reporting, and general internal correspondence. Several cross-cutting patterns stand out. First, many users are non-native English speakers writing in French, Portuguese, German, Spanish, Italian, Japanese, or other languages seeking both linguistic correction and translation. Second, a significant share of interactions involve diplomatically softening or formally elevating messages — navigating sensitive topics like performance feedback, employee departures, or escalation requests. Third, scientific and clinical content editing appears frequently: users refine sentences about drug mechanisms, adverse event documentation, protocol language, and clinical study updates. The AI functions almost entirely as a high-speed writing assistant — rephrasing, correcting, condensing, and restructuring user-supplied text.

- **Write material for artistic or commercial purposes:** In this IWA, AI assistance centers on producing polished original or near-original professional content rather than correcting existing text. Career-related content creation appears frequently: resume bullet points, LinkedIn-style professional headlines, development gap statements, and cover-letter-style narratives tailored to specific roles or hiring managers. A secondary cluster involves short-form creative or commercial writing: event promotion copy for oncology nursing webinars, team/program name brainstorming, and occasional lighter requests (limericks for internal celebrations). The AI's role blends ghostwriting with tone-setting — users provide the intent and context, and the AI produces polished output ready to send or present.
- **Present research or technical information:** This IWA captures interactions where AI helps employees make sense of, explain, or communicate complex scientific and technical content. A large share of cases involve summarizing or simplifying dense clinical or pharmaceutical material: users request plain-language explanations of clinical trial analysis sets, pharmacokinetic findings, statistical concepts, and drug class comparisons. Medical Q&A exchanges are common, with employees querying the AI on disease state definitions (multiple myeloma response criteria, SLE cognitive dysfunction, IBD terminology) and pharmaceutical stability zones. A second major thread is structuring or articulating research presentations: drafting email communications about study risks to leadership, writing background sections for AI governance or regulatory presentations, generating conference panel topics for ML/pharmacometrics applications, and producing multilingual translations of clinical operational updates. The AI also supports quick factual lookups — acronym definitions, SAP table structures, chemistry terminology — crossing into IT and data science domains.
- **Interpret language, cultural, or religious information for others:** Translation is the defining activity in this IWA, and the samples show an exceptionally broad multilingual footprint. Content ranges from single words and phrases to full paragraphs of regulatory text, clinical protocols, HR policies, internal announcements, and business communications. Subjects frequently involve pharmaceutical and clinical content — protocol amendments, pharmacovigilance terms, drug dosing instructions, adverse event narratives, and study status updates — as well as operational and administrative messages (HR orientation topics, invoice and procurement language, SAP guidance).
- **Respond to customer problems or inquiries:** This IWA encompasses a diverse mix of interactions where employees are looking for information, troubleshooting problems, or drafting responses to practical issues. Examples include employees seeking factual answers to operational or regulatory questions as well as drafting professional responses to customer or stakeholder-facing situations: polishing emails about device malfunctions, handling warranty follow-ups, and communicating QC discrepancies. The AI serves as both a quick reference resource and a writing aid for problem-resolution communications.
- **Prepare reports of operational or procedural activities:** In this IWA, AI is used extensively to help employees document, summarize, and communicate operational activities. The most common task type is converting raw or informal content into structured reports: users upload meeting transcripts, study status decks, or rough notes and ask for formal meeting minutes, action item summaries, or polished status updates. Subject matter includes clinical trial operations, procurement and finance, manufacturing, and organizational updates.
- **Provide information or assistance to the public:** This IWA spans a broad and eclectic range of information-seeking and communication-support interactions. A substantial portion involves employees asking the AI for factual explanations across pharmaceutical, clinical, regulatory, and technical domains: pharmacovigilance responsibilities, drug classifications, IBD disease concepts, oncology drug eligibility criteria, pharmaceutical stability zones, and global regulatory landscapes. IT and data science questions also appear frequently — employees ask about AWS services, Kubernetes architecture, code refactoring, and Excel/Power Query workflows.

- **Provide information to guests, clients, or customers:** In this IWA, AI is primarily used as a quick-access information and language resource for employees interacting with external parties — HCPs, patients, vendors, and partners. A large share of interactions involve translation, many in support of HCP-facing or patient-facing materials such as injection technique guidance, adverse event descriptions, drug dosing instructions, and clinical eligibility criteria. Employees also ask for factual pharmaceutical information to relay to physicians or patient support teams — drug-drug interaction concerns, device malfunction resolution steps, and reimbursement or access program details.
- **Explain technical details of products or services:** This IWA shows heavy usage across IT, data science, and pharmaceutical technical domains. On the technical side, employees ask for explanations and assistance with database schema design, data visualization configuration, automation flows, scripting, table structures, and Linux/Kubernetes administration. These interactions often involve troubleshooting specific errors or understanding system behavior. On the pharmaceutical and scientific side, employees request technical explanations of clinical concepts, regulatory procedures, and product-specific details. A recurring task is condensing dense technical content for presentation — polishing abstract text, converting policy language into plain English, and producing slide-ready summaries. The AI acts as both an explainer and a technical writing partner, helping employees bridge complex subject matter expertise and clear stakeholder communication.

From 2024 to 2025, AI task completion rates improved measurably across a variety of Intermediate Work Activities (IWAs). Evolved AI capabilities are not merely providing generic answers but completing discrete, well-scoped tasks with clear user intent. In practical terms, AI has become more capable of handling multi-step reasoning within a single conversation, enabling it to fully resolve technical troubleshooting requests or procedural questions rather than offering only partial guidance. Advances in model capabilities between 2024 and 2025, including improved instruction-following and longer effective context windows, are likely contributing factors [10]. Examples observed in the data shown in Figure 5 include asking AI to adjust technical formulas, navigate document management workflows, and align process changes to organizational frameworks. These tasks require AI to reason about structured systems, apply domain-specific knowledge, and produce outputs that are directly actionable.

*Figure 5: Biggest Completion Gains in IWAs, 2024 to 2025*

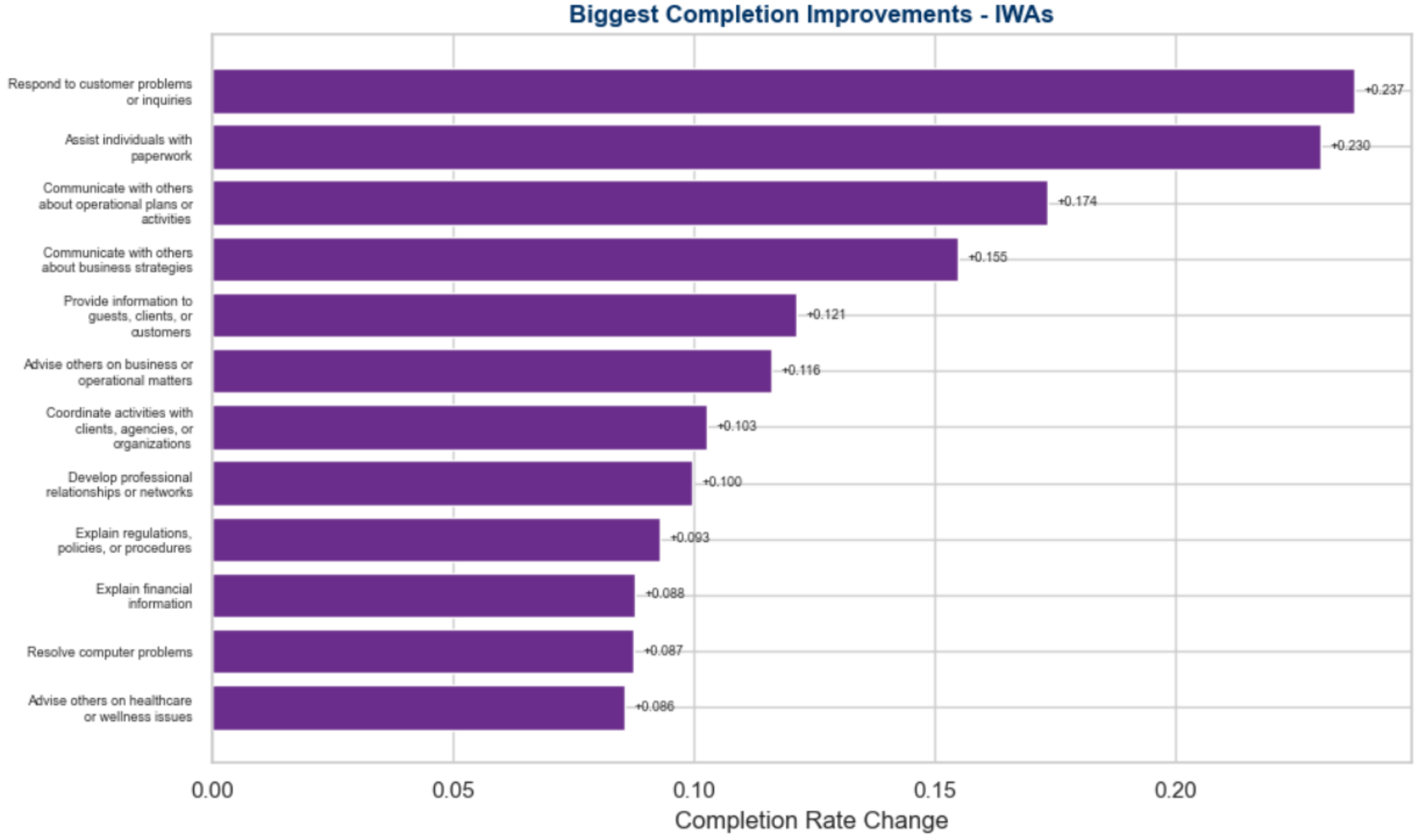

# AbbVie Intelligence 3.0 Release

In August 2025, the AbbVie Intelligence product underwent a significant product enhancement. The release of version 3.0 of the platform product distinct tools such as web search and literature search into a unified prompt-based flow. Instead of the user having to specify manually that they want to search the web, for example, version 3.0 employed an orchestration router to intelligently detect whether specific capabilities would be required for an accurate response. This boosted the prominence of search tools across the board, as many users defaulted to asking the bare LLM, and did not click on the UI elements required to enable those tools. Therefore, post-launch, we expected usage scenarios to begin to skew more toward up-to-date or real-time information gathering as users became more comfortable with those tools.

A pre-post quasi-experimental design was applied to AI applicability scores. Observations from the same set of occupations are compared across two-time windows:

1) Baseline period; data collected before the AbbVie Intelligence 3.0 product release.
2) Treatment period; data collected two months following the product release.

Because the same set of occupations is observed in both periods, a paired analysis design was used as the primary statistical approach. This approach controls for stable between-occupation differences and maximizes statistical power for detecting a true treatment effect. Of the 184 occupations in the pre-release dataset and 173 in the post-release dataset, 168 were present in both periods and were included in the paired analysis. The 16 occupations appearing only in the pre-release period and the 5 appearing only post-release were excluded from the primary paired analysis to preserve the matched structure of the study design.

*Table 1: Descriptive Statistics -- AI Applicability Scores, Pre- vs. Post-Release*

| Statistic | Pre-Release | Post-Release |
|---|---|---|
| N (occupations) | 168 | 168 |
| Mean | **0.2346** | **0.2580** |
| Median | 0.2436 | 0.2718 |
| Standard Deviation | 0.0993 | 0.1068 |
| Minimum | 0.0000 | 0.0000 |
| Maximum | 0.4611 | 0.5062 |
| Q1 (25th percentile) | 0.1702 | 0.1880 |
| Q3 (75th percentile) | 0.2994 | 0.3299 |

The mean AI Applicability score increased from 0.2346 (pre-release) to 0.2580 (post-release), representing an absolute change of +0.0235 and a relative improvement of +10.0%. The 95% confidence interval for the mean change is (+0.0215, +0.0254), indicating that the interval does not include zero and the observed increase is unlikely to be due to chance. The interquartile range shifted upward in the post-release period, suggesting broad-based improvement rather than gains concentrated at the tails of the distribution.

*Figure 6: AI Applicability Score Distribution, Pre- vs. Post-Release*

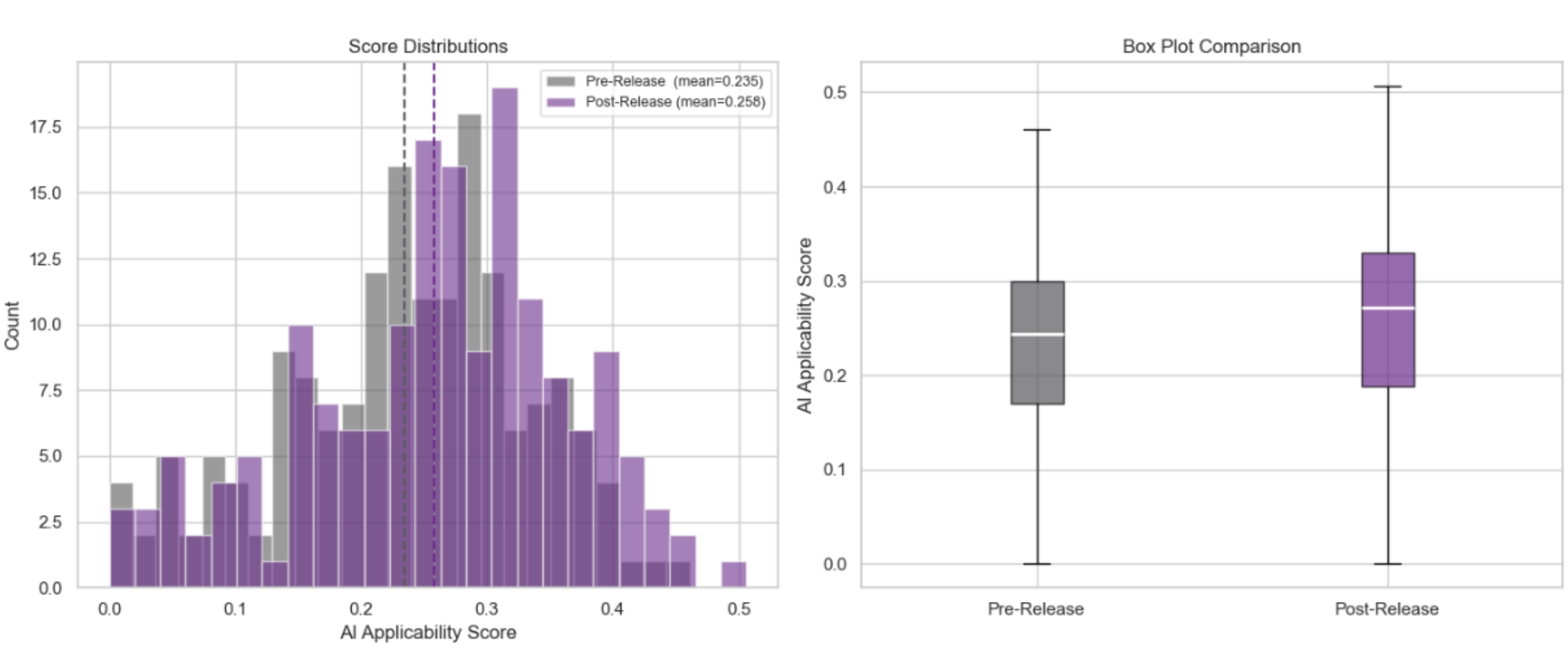

The paired t-test is the most appropriate primary test for this study design, as the same occupations are observed in both periods, enabling direct within-occupation comparison that eliminates confounding from stable occupation-level heterogeneity. The highly significant result ($t = 24.08$, $p < 0.001$) provides very strong statistical evidence that AI Applicability scores increased post-release.

The Wilcoxon signed-rank test (the non-parametric analogue to the paired t-test) corroborates this finding without relying on normality assumptions ($W = 13,860$, $p < 0.001$). The convergence of parametric and non-parametric results strengthens confidence in the conclusion.

Of the 168 matched occupations, 165 (98.2%) exhibited an increase in AI Applicability score following the release. Notable occupations that experienced the largest increases in AI Applicability score are shared in Figure 7.

*Figure 7: Top Occupations by AI Applicability Score Gain, Post-Release*

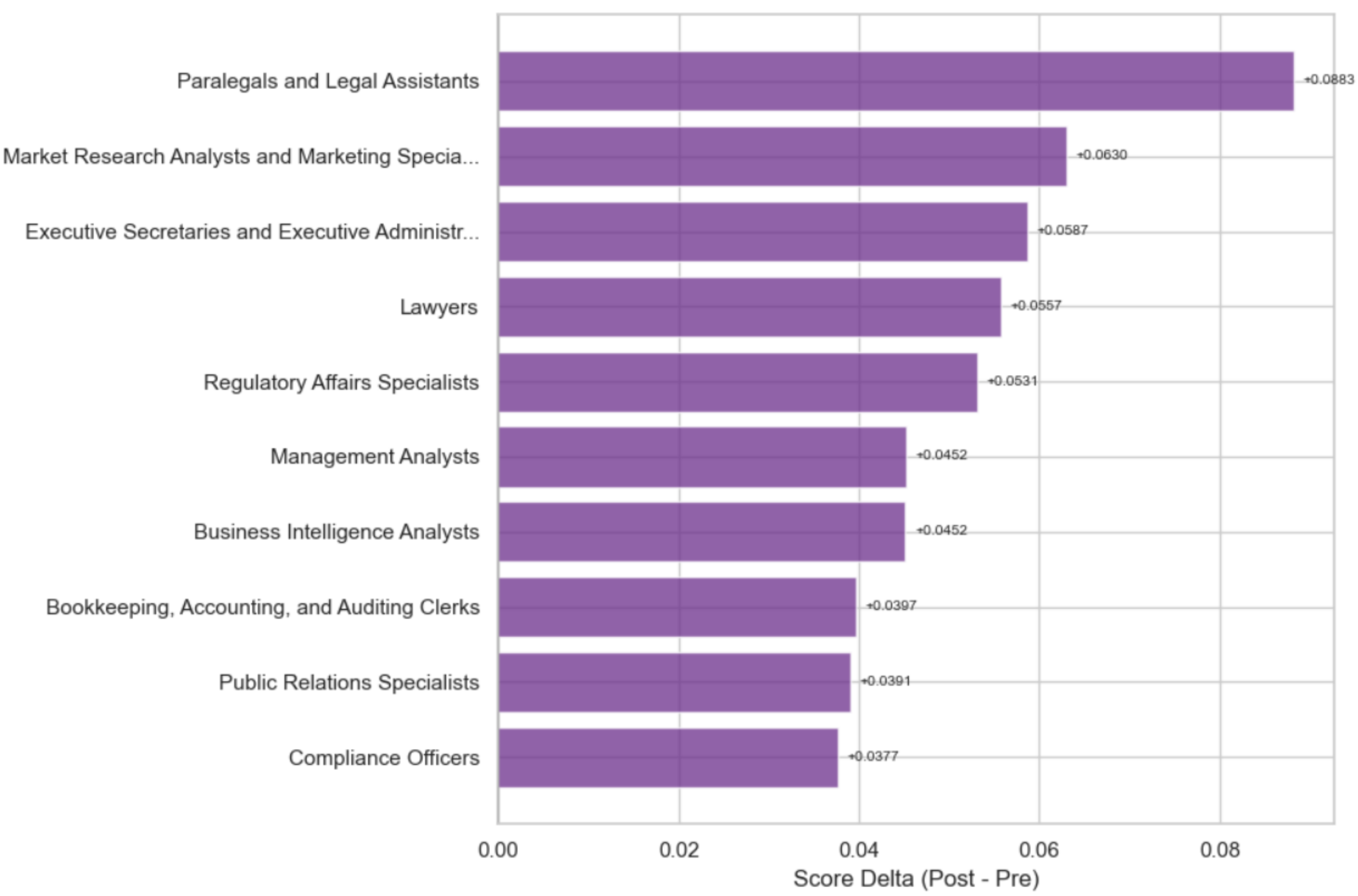

The pronounced score gains observed within the Legal function following the rollout of AbbVie Intelligence 3.0 are not incidental; they reflect a direct alignment between the platform's new capabilities and the needs this group had articulated well before the upgrade. In pre-launch surveys and working sessions, Legal professionals consistently identified advanced reasoning and the ability to configure AI behavior through custom system prompt language as the capabilities most critical to their work. These use cases had been constrained by earlier versions of the product. AbbVie Intelligence 3.0 introduced enhanced thinking models alongside flexible persona customization via custom system prompts. Post-launch feedback from the Legal cohort confirmed that these additions meaningfully expanded their practical use cases. The correlation between that expressed readiness and the subsequent score improvement offers a compelling illustration of how targeted capability development, informed by user input, can unlock engagement within specialized professional communities.

Beyond the occupation-level scores, Intermediate Work Activity (IWA) level changes were also examined to understand which specific task categories drove the observed improvements. Below are the IWAs with the largest jumps in activity.

- **Examine financial activities, operations, or systems:** A substantial portion involves financial reconciliation and verification — users checking whether invoices have been paid, investigating discrepancies between accounting systems (e.g., SAP vs. BlackLine, CN files vs. SAP accounts), verifying journal entry accuracy, resolving duplicate payments, and tracking down the originator of a financial transaction. In many of these cases, AI is being used as a communication or translation aid — helping users articulate reconciliation questions or write messages to colleagues — rather than directly resolving the discrepancy, since AI typically cannot access the underlying systems. A second major theme is financial accounting education and standards interpretation. Users ask AI to explain GAAP treatment of purchase price variances, the mechanics of CapEx and depreciation on financial statements, journal entries for accrued liabilities, the difference between operating earnings and division margin, how hospital ceilings and paybacks are accounted for in pharma, and how stock option compensation expense is calculated. A third significant theme is audit and compliance support — helping users structure internal audit approaches, identify risks, and draft communication. And finally, Payroll and HR-related financial inquiries round out the category: users ask AI about tax withholding discrepancies, bonus calculations, pay group errors in Workday, and compensation contribution disputes.

- **Manage budgets or finances:** Use cases span budget planning, allocation, tracking, and approval. Users engage AI to structure expense tracking systems (including Excel formulas for P&L management), split budgets across cost centers or floors, develop chargeback models for shared platform costs, generate performance goals that include budget monitoring metrics, plan event expenditures against available funds, forecast spend from historical data, and communicate budget decisions to stakeholders. AI's role in these conversations is often a combination of advisor, calculator, and writing assistant — helping users both do the financial work and communicate it clearly.
- **Mark materials or objects for identification:** AI is primarily being used to classify and label materials according to established taxonomies or organizational naming conventions. Users submit document titles or brief descriptions and ask AI to recommend the best-fitting classification — sometimes requesting ranked alternatives when the primary match is ambiguous.
- **Order medical tests or procedures:** AI is primarily assisting healthcare and clinical research professionals with the selection and sequencing of diagnostic tests. Conversations center on determining which tests are appropriate for a given clinical scenario — such as identifying the best lab panel for a patient with cardiovascular-related events, selecting between bone marrow aspirate versus core biopsy for confirming multiple myeloma response criteria, or deciding whether to collect tryptase and pharmacokinetic samples following an adverse event in a clinical trial.
- **Collect samples of products or materials:** AI is being used to guide the technical execution of sample collection procedures, helping users navigate the practical steps involved in collecting biological, chemical, and environmental samples. Conversations range from general procedural requests — such as step-by-step aseptic swab collection or general biological/environmental sampling methodology — to more specialized protocols. In these cases, AI functions as an on-demand technical reference, providing structured guidance on preparation, technique, and sample integrity during collection. AI also supports the logistical and analytical dimensions of sample collection (e.g. dilution math, target concentrations, etc.).
- **Direct security or safety activities or operations:** AI's role in these interactions is largely translational and advisory, helping users communicate clearly about security actions or navigate the relevant systems — such as deactivating system credentials for a terminated employee who may attempt facility entry, removing file and system access for specific users, and managing SAP authorization approval workflows. For operational safety, the use cases are more substantive. Users are seeking AI guidance on developing layered security strategies to prevent metal theft at construction sites, safely isolating a steam line using appropriate procedures or PPE, and implementing corrective actions following machinery accidents. In these conversations, AI provides structured safety frameworks — covering procedures, controls, and monitoring steps — that users can adapt for their specific environments.
- **Prepare medical equipment or work areas for use:** AI is widely used to provide step-by-step procedural guidance for preparing laboratory and clinical environments. A consistent theme across conversations is professionals asking AI to walk them through preparation protocols they may be encountering for the first time, performing infrequently, or executing under constraints. For these types of requests, AI provides structured, actionable instructions that reduce the cognitive burden of recalling detailed procedural steps from memory or documentation.

# AbbVie AI Summit

As AI tools become increasingly embedded in enterprise workflows, organizations face a critical question: do investments in AI learning and enablement programs translate into measurable improvements in how AI is applied to real work? The AbbVie AI Learning Summit, held on November 5, 2025, represented a deliberate, organization-wide effort to build AI awareness, fluency, and adoption.

The AI Learning Summit marked an important step toward developing AI leadership and innovation by focusing on prompt engineering and responsible AI use. Participants learned how to effectively communicate with AI using prompts grounded in business context, enabling them to unlock AI's value within their specific domains. The sessions covered multiple prompt-creation techniques, personalization of AI to reflect individual voice and improve efficiency, and how to "talk to a file" by using AI with personal documents such as SOPs, procedures, project plans, and research materials.

Functional leads further demonstrated how AI can be applied to job-specific tasks by sharing practical, real-world prompts that support daily work. Attendees were encouraged to practice these skills using the provided step-by-step guide and continue their learning through the AI Learning Hub, an internal AI learning platform which houses session recordings and upcoming hands-on learning opportunities.

Statistical analysis of AI applicability scores, pre-summit versus post-summit, can shed some light on the impact of investments in AI education. Of 173 pre-summit and 171 post-summit occupations, 163 appeared in both periods and were included in the paired analysis. Below are summary statistics of the 163 matched occupations in both periods.

*Table 2: AI Applicability Scores, Pre- vs. Post-Summit*

| Statistic | Pre-Summit | Post-Summit |
|---|---|---|
| N (occupations) | 163 | 163 |
| Mean | **0.2596** | **0.2770** |
| Median | 0.2725 | 0.2901 |
| Standard Deviation | 0.1056 | 0.1115 |
| Minimum | 0.0000 | 0.0000 |
| Maximum | 0.5062 | 0.5225 |
| Q1 (25th percentile) | 0.1947 | 0.2118 |
| Q3 (75th percentile) | 0.3300 | 0.3515 |

Pre- versus post-summit comparison shows a modest rise in mean AI Applicability score, from 0.2596 to 0.2770. Once again, there is broad improvement across the distribution, as indicated by upward shifts in the interquartile range. The paired t-test ($t = 11.27$, $p < 0.001$) and Wilcoxon signed-rank test ($W = 12{,}471$, $p < 0.001$) both provide highly significant evidence of a post-summit increase.

*Figure 8: AI Applicability Score Distributions, Pre- vs. Post- Summit*

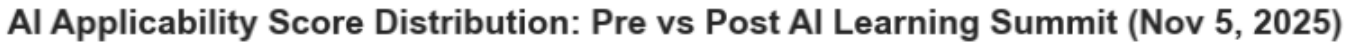

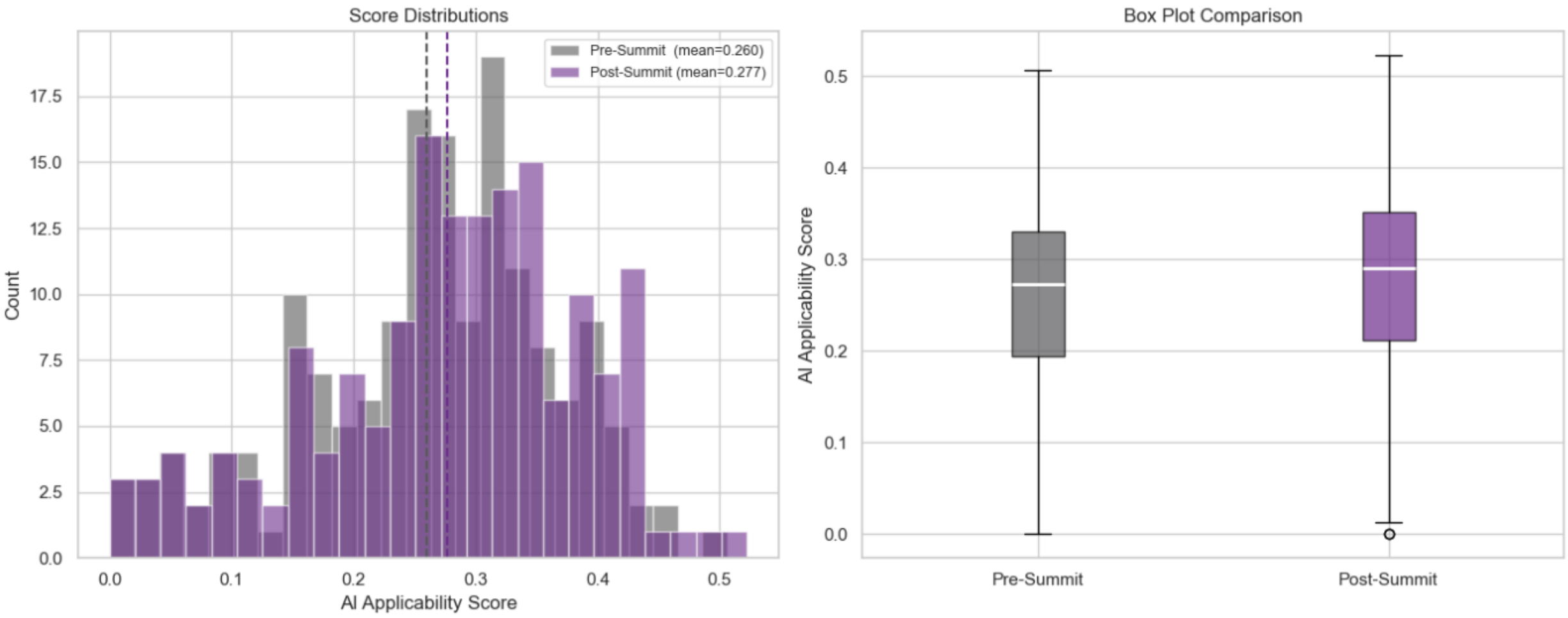

These results provide strong evidence that the AI Learning Summit meaningfully contributed to expanding AI's applicability across a broad range of AbbVie occupational categories, consistent with the summit's goal of accelerating AI adoption and capability development.

*Figure 9: Top 10 Occupations by AI Applicability Score Gain, Post-Summit*

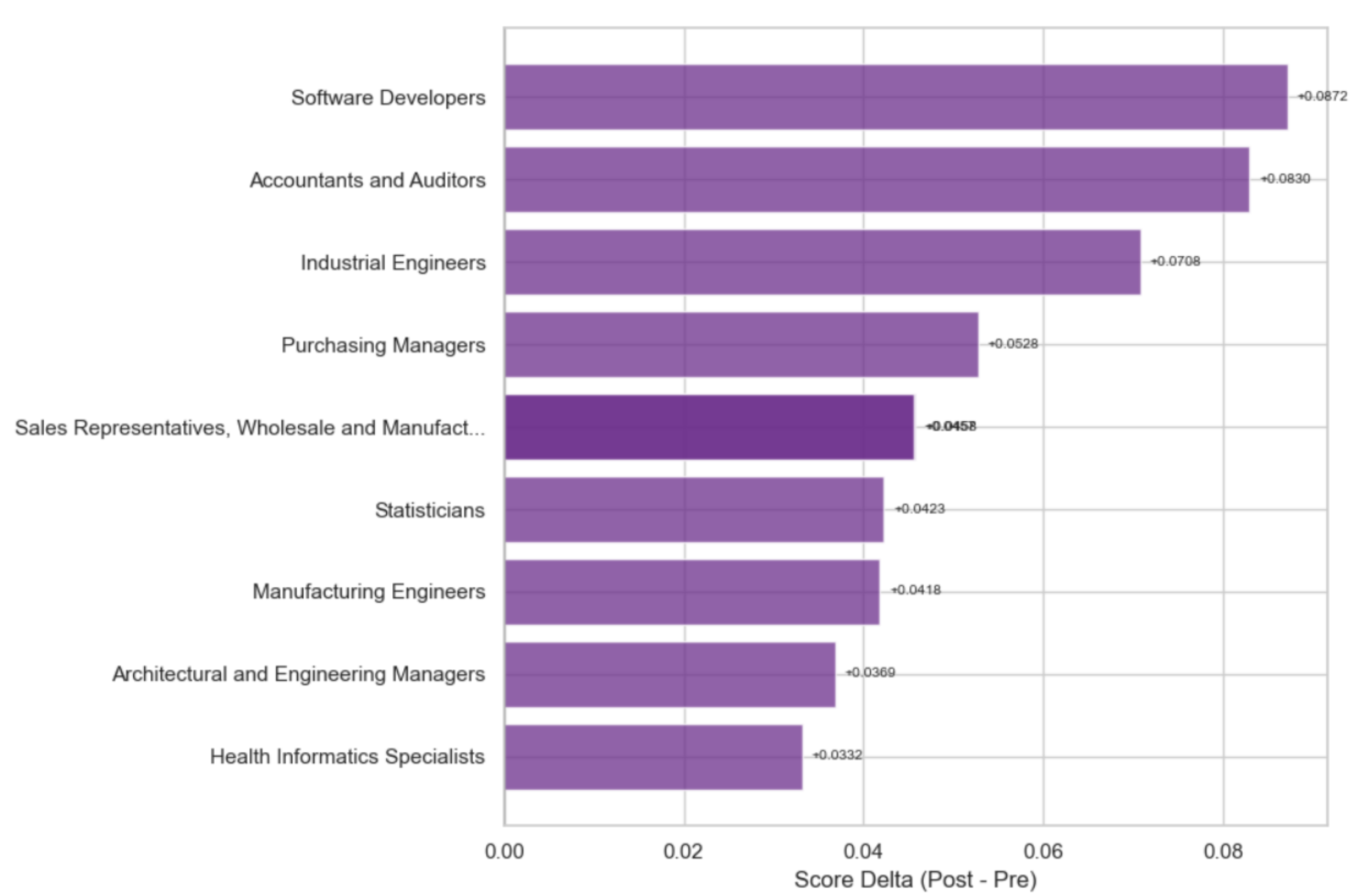

# Overall Adoption and Estimated Value Impact

The primary focus of this study on the specific IWAs and occupations must also be framed within a context of the broader increase of general product adoption and usage volume. Additionally, an estimation of the value impact to the business is also an important aspect of framing the applicability and impact of AI tools within the pharmaceutical workplace.

While a primary driver of AI adoption early on was a growing sense of FOMO (fear of missing out), the 2025 KPMG American Worker Survey illustrated a shift to a sense of FOBO (fear of being obsolete). This survey aligns with the consensus of AbbVie employees post the AI Summit event. While demonstrating the accessibility of the AbbVie Intelligence product and its improved capabilities through hands-on and practical activities, employees left the session feeling empowered to tackle more complex work activities. Not only are there significant increases in the number of employees choosing to engage in AI-assisted work after these sessions, those using the product also demonstrate increased individual utilization. While some of this may be attributed to general popular awareness, it is important to note that almost every statistically significant increase of utilization over and above the baseline growth through two years of product availability can be tied directly to internal product communications, audience targeted presentations, and key enablement events/activities.

*Figure 10: AbbVie Intelligence Usage and Adoption, 2024-2025*

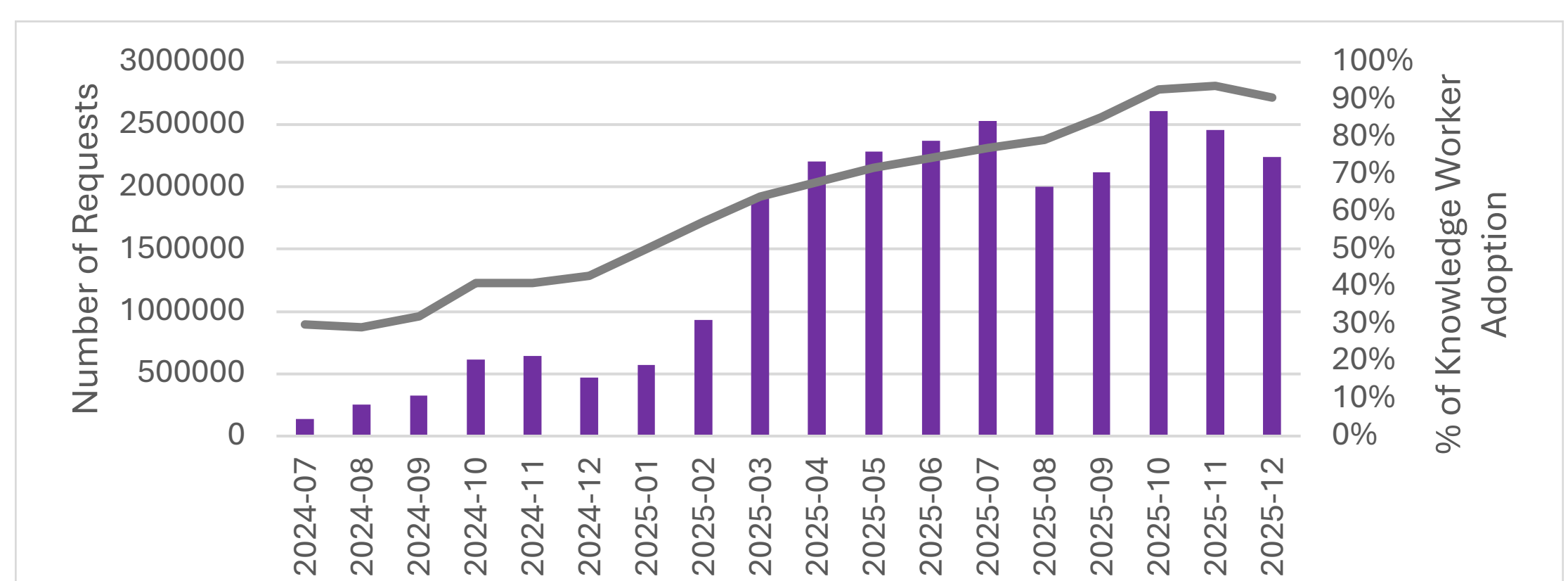

While adoption is consistently growing and usage continues at over 2M requests per month on average, the financial impact of this investment is more difficult to directly calculate in its entirety as there are many indirect returns from this growing use and engagement.

- **Reduced Iterations** – providing examples for higher quality prompting and clearer direction on input content and context drives a proportionate improvement in the quality of outputs. These higher quality outputs drive down the number of iterations and number of human edits and validation required on the output side of the task.
- **Speed Improvements** – the speed improvements of modern tools have a direct impact on the time individual tasks can require for successful completion. However, the time compression impact is difficult to quantify without larger efforts to characterize baseline cost-vs-time characteristics of the aggregate work process versus individual tasks.
- **Output Precision** – the improvements of input quality and modern models/tools also drive to a higher precision in the quality of outputs. Measuring the impact of more precise outputs such as better classifications or higher fidelity inferences can demonstrate significant follow-on impacts by reducing the complexity of human edits and validation required on the output side of the task.

Even with these complexities, we can choose an arbitrary but informed estimate of time saved per IWA and use that to extrapolate a low/high range of potential ROI to the enterprise.  In Table 3 we present just such an estimate demonstrating a substantial estimate of time savings.

*Table 3: Estimated Impact of IWA Time Saved, 2024 and 2025*

| IWA | Estimate Time Saved per Task | 2024 Impact (hrs) | 2025 Impact (hrs) |
|---|---|---|---|
| Edit written materials or documents | 10-15 minutes | 7494 | 11464 |
| Explain technical details of products or services | 15-30 minutes | 5520 | 5922 |
| Interpret language, cultural, or religious information for others | 10-15 minutes | 184 | 4611 |
| Prepare informational or instructional materials | 20-40 minutes | 3503 | 31210 |
| Prepare reports of operational or procedural activities | 20-40 minutes | 1807 | 10423 |
| Present research or technical information | 15-30 minutes | 2439 | 10987 |
| Provide information or assistance to the public | 10-15 minutes | 4715 | 5793 |
| Provide information to guests, clients, or customers | 5-10 minutes | 708 | 2570 |
| Respond to customer problems or inquiries | 10-15 minutes | 72 | 5695 |
| Write material for artistic or commercial purposes | 10-15 minutes | 836 | 6168 |
| **Total Estimate of Impact** | | **27278 (19959-34598)** | **94842 (67725-121959)** |

An important note about communicating this impact and savings estimate is that it is inappropriate to directly convert these time savings into a commensurate reduction in employee headcount.  As these impacts and savings are distributed over different IWAs and occupations, it is unlikely to result in a 1-1 conversion/savings.  Table 4 demonstrates how even though the estimate of time saved per task is consistent amongst the top occupations, the differential impact per role becomes a function of the volume of activity and distribution of activity amongst higher/lower return IWAs.

*Table 4: Estimate of Time Saved per Task by Occupation, 2024 to 2025*

| Occupations | Estimate Time Saved per Task | 2024 Impact (hrs/requests) | 2025 Impact (hrs/requests) |
|---|---|---|---|
| Business Intelligence Analysts | 12.6 minutes | 668 / 2532 | 1333 / 8475 |
| Clinical Research Coordinators | 12.6 minutes | 7926 / 30087 | 16525 / 106198 |
| Compliance Managers | 12.7 minutes | 2568 / 9924 | 4774 / 28782 |
| Computer Systems Engineers/Architects | 14.7 minutes | 2023 / 6935 | 2995 / 15182 |
| Human Resources Specialists | 12.4 minutes | 995 / 3764 | 2245 / 15023 |
| Medical and Health Services Managers | 13.2 minutes | 9809 / 36348 | 17575 / 103829 |
| Regulatory Affairs Specialists | 13.0 minutes | 1141 / 4289 | 1923 / 11444 |
| Statisticians | 14.4 minutes | 658 / 2239 | 1019 / 5495 |
| Training and Development Managers | 13.7 minutes | 1042 / 3670 | 1887 / 10993 |
| Training and Development Specialists | 13.5 minutes | 448 / 1569 | 884 / 5401 |
| **Total Estimate of Impact** | | **27278 / 101357 (19959-34598)** | **94842 / 310822 (67725-121959)** |

This is the biggest challenge in justification of the value of impact for general employee productivity tools delivered by AI. The implication of these numbers is that while on aggregate the value delivery to the enterprise is measured in tens to hundreds of thousands of hours, the individual impact to employees on average might only be a few hours of time total with a high level of variability based on occupation and individual adoption level.

# Discussion

This study provides quantitative evidence that AbbVie Intelligence has become a broadly applicable and continuously improving tool across the enterprise workforce. Three convergent lines of analysis—longitudinal year-over-year trends, a product release pre-post comparison, and an AI education intervention evaluation—collectively establish that both technological investments and organizational capability programs drive meaningful, statistically significant gains in AI applicability.

**Longitudinal Growth: 2024 to 2025**

The year-over-year comparison reveals a fundamental shift in how AI is used across AbbVie. The observed growth was not concentrated in a narrow set of roles. Non-clinical and technology occupations showed accelerating adoption, indicating that AI usage is expanding beyond early-adopter clinical and research functions into operational and corporate domains. The most improved roles reflect both the increasing sophistication of AI capabilities in technical domains and a growing organizational comfort with applying AI to high-value, specialized work.

**Impact of the AbbVie Intelligence 3.0 Release**

The AbbVie Intelligence 3.0 release in August 2025 produced a +10.0% increase in mean AI Applicability Scores (0.2346 to 0.2580). Both the paired t-test ($t = 24.08$, $p < 0.001$) and the Wilcoxon signed-rank test ($W = 13{,}860$, $p < 0.001$) confirmed a highly significant, robust effect. Critically, 165 of 168 matched occupations (98.2%) showed score increases, demonstrating that improvement was enterprise-wide rather than driven by a small number of outlier roles. The platform's unification of web search and literature search into a seamless, prompt-based workflow lowered friction for information-intensive tasks, allowing employees to accomplish more complex, multi-step work within single AI interactions.

The IWAs with the largest post-release gains—examining financial activities, managing budgets, ordering medical tests, and directing security operations—suggest that the new capabilities particularly benefited roles requiring current, contextual, or domain-specific information. The web search integration allowed employees to use AI as a genuine research and synthesis tool rather than a static knowledge repository, expanding applicability into previously underserved work categories.

**Impact of the AI Learning Summit**

The AbbVie AI Learning Summit produced a +6.68% increase in mean AI Applicability Scores (0.2596 to 0.2770), supported by strong statistical evidence ($t = 11.27$, $p < 0.001$; Wilcoxon signed-rank test: $p < 0.001$). This result is noteworthy for two reasons. First, it confirms that enterprise AI education programs can produce measurable, quantifiable changes in how employees interact with AI—an outcome that is often assumed but rarely demonstrated empirically. Second, the magnitude of the educational effect (+6.68%) is comparable to—though somewhat smaller than—the product release effect (+10.0%), establishing that human capability development is a complementary and independently powerful lever alongside technological improvement.

The summit's focus on prompt engineering and functional use-case demonstrations appears to have catalyzed more purposeful, task-aligned AI usage. Employees who learned to articulate their work context in structured prompts were better positioned to engage AI with clearly scoped goals that map cleanly onto scoreable IWAs, increasing both task completion rates and scope assessments in the applicability formula.

**Comparative Interpretation**

Both the AI summit and product release produced statistically significant increases with large paired effect sizes. The product release generated a somewhat larger absolute mean gain (+10.0% vs +6.68%), which is expected given that platform enhancements directly expand AI capability. The summit's +6.68% gain -- achieved through a learning event rather than a platform update -- is nevertheless substantial and demonstrates that human enablement investments can drive meaningful AI applicability improvements independent of product changes.

*Table 5: Mean Change Comparison, Product Release and AI Summit Impact*

| Metric | Product Release | AI Learning Summit |
|---|---|---|
| Pre-period mean score | 0.2346 | 0.2596 |
| Post-period mean score | 0.2580 | 0.2770 |
| Mean change | +0.0235 (+10.0%) | +0.0173 (+6.68%) |
| Occupations improved | 165 / 168 (98.2%) | 150 / 163 (92.0%) |

The slightly larger effect of the platform release relative to the AI Learning Summit is consistent with theoretical expectations: direct capability expansion creates new applicability possibilities for all users simultaneously, while education programs require individual behavior change and have uneven reach. However, the two mechanisms are not substitutes. The convergence of both in the 2025 period likely produced compounding benefits, as employees with stronger AI skills were better equipped to leverage the new capabilities introduced in AbbVie Intelligence 3.0.

**Limitations**

Several limitations should be considered when interpreting these findings. First, the pre-post study design, while strengthened by paired structure and non-parametric validation, cannot fully rule out concurrent confounders such as seasonal variation in AI usage patterns or simultaneous changes in workforce composition. Second, the AI Applicability Score reflects the scope and completion of AI-assisted activities as observed in conversation data; it does not directly measure downstream productivity outcomes such as time savings, error reduction, or quality improvement. Third, occupational classifications rely on O*NET IWA mappings that may not perfectly capture the specificity of AbbVie's internal job functions. Fourth, the LLM used to summarize and label the conversation data differed between 2024 and 2025. Data analyzed for 2024 was processed using GPT-4o, while 2025 labeling used GPT-5.1. This may introduce label and summary drift, making year-over-year comparisons less reliable because observed differences may reflect model changes rather than true changes in the underlying data. Finally, because conversations were classified using LLM-based classifiers, this introduced a degree of measurement error that is difficult to fully quantify, though data privacy guardrails ensured all processing operated on deidentified data.

**Implications**

These findings carry practical implications for AI investment strategy, workforce planning, and change management at AbbVie and analogous organizations. The broad-based score improvements across nearly all occupations suggest that AI applicability is not a niche phenomenon concentrated in a handful of technical roles, but a pervasive shift affecting the full knowledge-work workforce. Organizations should therefore invest in AI infrastructure and enablement programs as general workforce capabilities rather than narrow functional tools.

The demonstrated impact of AI education programs provides an empirical basis for continued investment in structured AI learning initiatives. Longitudinal tracking of AI Applicability Scores against business outcomes—productivity metrics, time-to-decision, quality indicators—represents a critical next step toward fully characterizing the return on investment from enterprise AI deployments.

# Appendix

## A.1 Microsoft's AI Applicability Formulas

*Figure 11: AI Applicability Score Formula*

$$a_i^{user} = \sum_{j \in \mathbf{IWA}_i} w_{ij} 1[f_j \geq 0.0005] C_j S_j^{user}$$

$$a_i^{AI} = \sum_{j \in \mathbf{IWA}_i} w_{ij} 1[f_j \geq 0.0005] C_j S_j^{AI}$$

$$a_i = \frac{a_i^{user} + a_i^{AI}}{2}$$

- $i$ = occupation (O*NET SOC code)
- $j$ = Intermediate Work Activity (IWA)
- $w_{ij}$ = importance & relevance-weighted fraction of work in occupation $i$ composed of IWA $j$
- $f_j$ = activity share (frequency threshold: 0.05%)
- $C_j$ = task completion rate for IWA $j$
- $S_j^{user}$ = scope_assistance (fraction where user-side scope is "moderate" or higher)

Example calculation based on data table below produces the following scores:

- Applicability Score User Goal: 0.100373913
- Applicability Score AI Action: 0.086988806
- AI Applicability Score: 0.093681359

| $i$ | $\mathrm{IWA}_i$ | $w_{ij}$ | $f_j$ | *covered* | $C_j$ | $S^{USER}$ | $S^{AI}$ | $a^{USER}$ | $a^{AI}$ |
|---|---|---|---|---|---|---|---|---|---|
| 11-1011.00 | Advise others on business or operational matters | 0.015051 | 0.020984738 | TRUE | 0.886356354 | 0.655443 | 0.547114852 | 0.008744028 | 0.007298866 |
| 11-1011.00 | Manage budgets or finances | 0.017547 | 0.002127139 | TRUE | 0.773967552 | 0.369223 | 0.69100295 | 0.005014422 | 0.009384514 |
| 11-1011.00 | Analyze data to improve operations | 0.067448 | 0.00184155 | TRUE | 0.772827892 | 0.510515 | 0.589374654 | 0.026611102 | 0.03072176 |
| 11-1011.00 | Implement procedures or processes | 0.017547 | 0.001609349 | TRUE | 0.719213313 | 0.254463 | 0.666868381 | 0.003211378 | 0.008416025 |
| 11-1011.00 | Assign work to others | 0.031981 | 7.29E-05 | FALSE | 0.62890625 | 0.382813 | 0.28125 | 0 | 0 |
| 11-1011.00 | Draft legislation or regulations | 0.081461 | 0.031572867 | TRUE | 0.922981234 | 0.755355 | 0.414534167 | 0.056792983 | 0.031167641 |

# A.2 Classification Prompts

The following are the prompts used to classify work activities as referenced from the Microsoft AI Applicability paper.

## General Prompt:

```
"""<|Instruction|> \n
# Task overview \n
You will be given a conversation between a User and an AI chatbot. \n
You have two primary goals: \n
(1) summarize the main goal that the user is trying to accomplish in the style of an O*NET Intermediate Work Activity (IWA). \n
(2) summarize the action that the bot is performing in the conversation in the style of the O*NET Intermediate Work Activities provided in this prompt. For
example, if the user asks for help with a computer issue and the bot provides suggestions to resolve the issue, the users IWA is "Resolve computer problems" and
the bots IWA is "Advise others on the design or use of technologies" \n
\n
\n
Sometimes, the user intent and bot action may be the same. For instance, if the user asks the bot to spellcheck a research paper and the bot corrects a few
misspelled words, the users IWA is "Edit written materials or documents" and the bots IWA is also "Edit written materials or documents" \n
\n
\n
For both the user and bot IWA summaries, you will generate several variations of the summary to capture the same intent using different wordings. \n
To aid your analysis, you will also summarize the conversation. \n
Finally, you will also determine whether the User is a student trying to do homework. \n
# Task details \n
Your task is to fill out the following fields: \n
summary: Summarize Users queries in 3 sentences or fewer in **English**. \n
user_iwa: Summarize the task the user is trying to accomplish using verbatim phrasing from the Intermediate Work Activities (IWA) provided in this prompt. Ensure
that the summary accurately describes the goal of the User as directly evidenced in the conversation. Ensure that the summary matches the level of generality of an
O*NET IWA: it should general enough to be an activity performed in a large number of occupations across multiple job families, but specific enough to capture the
essence of the Users goal. Provide exactly one succinct IWA-style summary. \n
user_iwa_variations: Generate a maximum of 4 IWAs performed by the user with verbatim IWA phrasing. \n
bot_iwa: Summarize the task that the bot is performing using verbatim phrasing from the Intermediate Work Activities (IWA) provided in this prompt. Ensure that the
summary matches the level of generality of an O*NET IWA: it should general enough to be an activity performed in a large number of occupations across multiple job
families, but specific enough to capture the essence of the bots actions. Provide exactly one succinct IWA-style summary. \n
bot_iwa_variations: Generate a maximum of 4 IWAs performed by the bot using verbatim IWA phrasing. \n
\n
\n
# Hints \n Provide your answers in **English** using the given structured output format. \n
<|end Instruction|> \n
\n
\n
<|Intermediate Work Activity (IWAs)|>

\n

    """
```

# Classify User Prompt:

```
"""<|Instruction|>
# Task overview
You will be given a conversation between a User and an AI chatbot as well as a summary of the conversation and a list
,→ of Candidate Intermediate Work Activity (IWA) statements from O*NET.
The IWAs will be numbered with numerical IDs to help you reference them in your responses.
Your primary task is to determine for each of the Candidate IWAs whether the user is trying to perform that IWA,
according to the meaning of the IWA in the context of O*NET. The conversation must provide direct evidence that
the user is themself trying to accomplish the IWA.
,→
,→
For example, a user asking for tech support does not match a IWA about providing tech support, but does match a IWA
,→ about resolving technical issues.
As another example, a user seeking information about a product does not match a IWA about providing product
,→ information, but does match a IWA about researching product information.
Additionally, you will determine the level of assistance that the bot provides to the user in the conversation for
,→ each matching IWA.
# Task details
Your reply to iwa_analyses should be a list of UserIWAAnalysis objects, one for each Candidate IWA in the order
below. For each Candidate IWA, you will analyze the user's intent relative to that IWA and fill out the fields of
UserIWAAnalysis as follows:
,→
,→
iwa (str): Copy the current Candidate IWA verbatim into this field. All of the following fields will be based on this
,→ IWA.
iwa_explanation (str): Explain in one sentence what the IWA means in the context of O*NET and what kinds of
,→ occupations perform this IWA.
is_match_explanation (str): Explain in one sentence whether the user is seeking to perform an activity described by
the IWA, according to the meaning of the IWA in O*NET. To be considered a match, the user's intent must be to
perform the action themselves, so if the IWA mentions or implies assisting clients or customers, for instance,
there must be evidence in their query that the user is seeking to assist a client or customer.
,→
,→
,→
is_match (bool): Based on your explanation, provide the label True if the user is seeking to perform an activity
described by the IWA, according to the meaning of the IWA in O*NET, and False otherwise. To be considered a
match, the user's intent must be to perform the action themselves.
,→
,→
assistance_level_explanation (str): Consider the full scope of the work performed under this IWA across all
occupations. What fraction of this work can the bot assist users with by applying only the capability it
demonstrates in this conversation? Pay careful attention to the fact that the IWA might encompass many more
subtasks than represented in this conversation. Explain in one sentence, or reply N/A if the IWA does not match
the user's intent (i.e., when is_match is False).
,→
,→
,→
,→
assistance_level (IWAAssistanceLevel): Based on your explanation, label the bot's capability to assist with the IWA
,→ using the IWAAssistanceLevel enum, which has the following options:
- none: The user is not seeking to perform the IWA, or the conversation does not indicate that the bot is capable of
,→ assisting with the IWA.
- minimal: With this demonstrated capability, the bot can assist with a minimal portion of the work in the IWA.
- limited: With this demonstrated capability, the bot can assist with a limited portion of the work in the IWA.
- moderate: With this demonstrated capability, the bot can assist with a moderate portion of the work in the IWA.
- significant: With this demonstrated capability, the bot can assist with a significant portion of the work in the
,→ IWA.
- complete: With this demonstrated capability, the bot can assist with all of the work in the IWA.
    """
```

# Classify Bot Prompt:

```
"""<|Instruction|>
# Task overview
You will be given a conversation between a User and an AI chatbot as well as a summary of the conversation and a list
,→ of Candidate Intermediate Work Activity (IWA) statements from O*NET.
The IWAs will be numbered with numerical IDs to help you reference them in your responses.
Your task is to determine for each of the Candidate IWAs whether the bot is performing that IWA in the conversation,
,→ based on the meaning of the IWA in the context of O*NET.
For example, if the user asks for help with a computer issue and the bot provides suggestions to resolve the issue,
,→ this matches an IWA about providing tech support, as that is the task that the bot is performing.
However, if the user asks the bot to spellcheck a research paper and the bot corrects a few misspelled words, this
does not match an IWA about writing research papers: while the **user's** overarching goal may be writing
research papers, that does not match the **bot's** task in the conversation.
,→
,→
Additionally, you will assess whether this conversation demonstrates the bot's ability to automate each matching IWA
,→ in the conversation.
# Task details
Your reply to iwa_analyses should be a list of BotIWAAnalysis objects, one for each candidate IWA in the order below.
For each candidate IWA, you will analyze the bot's actions relative to that IWA and fill out the fields of
BotIWAAnalysis as follows:
,→
,→
iwa (str): Copy the current Candidate IWA verbatim into this field. All of the following fields will be based on this
,→ IWA.
iwa_explanation (str): Explain in one sentence what the IWA means in the context of O*NET and what kinds of
,→ occupations perform this IWA.
is_match_explanation (str): Explain in one sentence whether the action that the bot is performing in the conversation
,→ is an example of a work activity described by the IWA, given the meaning of the IWA in the context of O*NET.
is_match (bool): Based on your explanation, provide the label True if the action that the bot is performing in the
conversation is an example of a work activity described by the IWA, given the meaning of the IWA in the context
of O*NET, and False otherwise.
,→
,→
automation_level_explanation (str): Consider the full scope of the work performed under this IWA across all
occupations. What fraction of this work can the bot perform by applying only the capability it demonstrates in
this conversation? Pay careful attention to the fact that the IWA might encompass many more subtasks than
represented in this conversation. Explain in one sentence, or reply N/A if the IWA does not match the bot's
action (i.e., when is_match is False).
,→
,→
,→
,→
automation_level (IWAAutomationLevel): Based on your explanation, label the bot's capability to perform the IWA using
,→ the IWAAutomationLevel enum, which has the following options:
- none: The bot does not perform the IWA, or the conversation does not indicate that the bot is capable of performing
,→ the IWA.
- minimal: With this demonstrated capability, the bot can perform a minimal portion of the work in the IWA.
- limited: With this demonstrated capability, the bot can perform a limited portion of the work in the IWA.
- moderate: With this demonstrated capability, the bot can perform a moderate portion of the work in the IWA.
- significant: With this demonstrated capability, the bot can perform a significant portion of the work in the IWA.
- complete: With this demonstrated capability, the bot can perform all of the work in the IWA.

    """
```

# Task Completion:

```
"""<|Instruction|>
# Task overview
You will be given a conversation between a User and an AI chatbot.
You will summarize the main task that the user is trying to accomplish in the conversation.
You will also determine whether the AI chatbot is able to complete the task, and if so, whether it reduced the time
,→ it takes to complete the task with equivalent quality by at least half.
# Task details
Your task is to fill out the following fields:
task_summary: Summarize the task the User is trying to accomplish in **English**.
completed_explanation: Explain in one sentence whether the AI chatbot is able to complete the User's task, based on
,→ the conversation.
completed: Based on your explanation, provide one of the following labels:
- not_complete: The AI chatbot did not make substantive progress towards completing the User's task.
- partially_complete: The AI chatbot made progress towards completing the User's task, but did not complete it.
- complete: The AI chatbot completed the User's task.
speedup_50pct_explanation: Explain in one sentence whether the AI chatbot reduced the time it takes to complete the
,→ task with equivalent quality by at least half. This includes tasks that can be reduced to:
- Writing and transforming text and code according to complex instructions,
- Providing edits to existing text or code following specifications,
- Writing code that can help perform a task that used to be done by hand,
- Translating text between languages,
- Summarizing medium-length documents,
- Providing feedback on documents,
- Answering questions about a document, or
- Generating questions a user might want to ask about a document.
Assume the user is a worker with an average level of expertise in their role trying to complete the given task.
speedup_50pct: Based on your explanation, provide the label True if the AI chatbot reduced the time it takes to
,→ complete the task with equivalent quality by at least half, and False otherwise.
# Hints
Provide your answers in **English** using the given structured output format.
<|end Instruction|>
    """
```